\documentclass[aps,twocolumn,pre,showpacs]{revtex4}

\usepackage{amsfonts}
\usepackage{amsmath}
\usepackage{amssymb}
\usepackage{graphicx}
\usepackage{color}

\begin{document}

\title{Linear and nonlinear rheology of dense emulsions: Identifying the glass and jamming regimes}

\author{Frank Scheffold$^{1,\ast}$, Fr\'ed\'eric Cardinaux$^{1}$ and Thomas G. Mason $^{2,3}$}

\address{$^{(1)}$Physics Department, University of Fribourg, Chemin de Mus\'{e}e 3, 1700 Fribourg, Switzerland}
\address{$^{(2)}$ Department of Chemistry and Biochemistry, University of California Los Angeles, Los Angeles, California 90095, USA }
\address{$^{(3)}$Department of Physics $\&$ Astronomy, University of California Los Angeles, Los Angeles, California 90095, USA}

\date{\today}

\begin{abstract}
We discuss the linear and non-linear rheology of concentrated (sub)microscale emulsions, amorphous disordered solids composed of repulsive and deformable soft colloidal spheres. Based on recent results from simulation and theory, we derive quantitative predictions for the dependences of the elastic shear modulus and the yield stress on the droplet volume fraction. The remarkable agreement with experiments we observe supports the scenario that the repulsive glass and the jammed state can be clearly identified in the rheology of soft spheres at finite temperature while crossing continuously from a liquid to a highly compressed yet disordered solid. 
\end{abstract}

\maketitle

\begin{figure}[htb]
\centerline{\includegraphics[width=8cm]{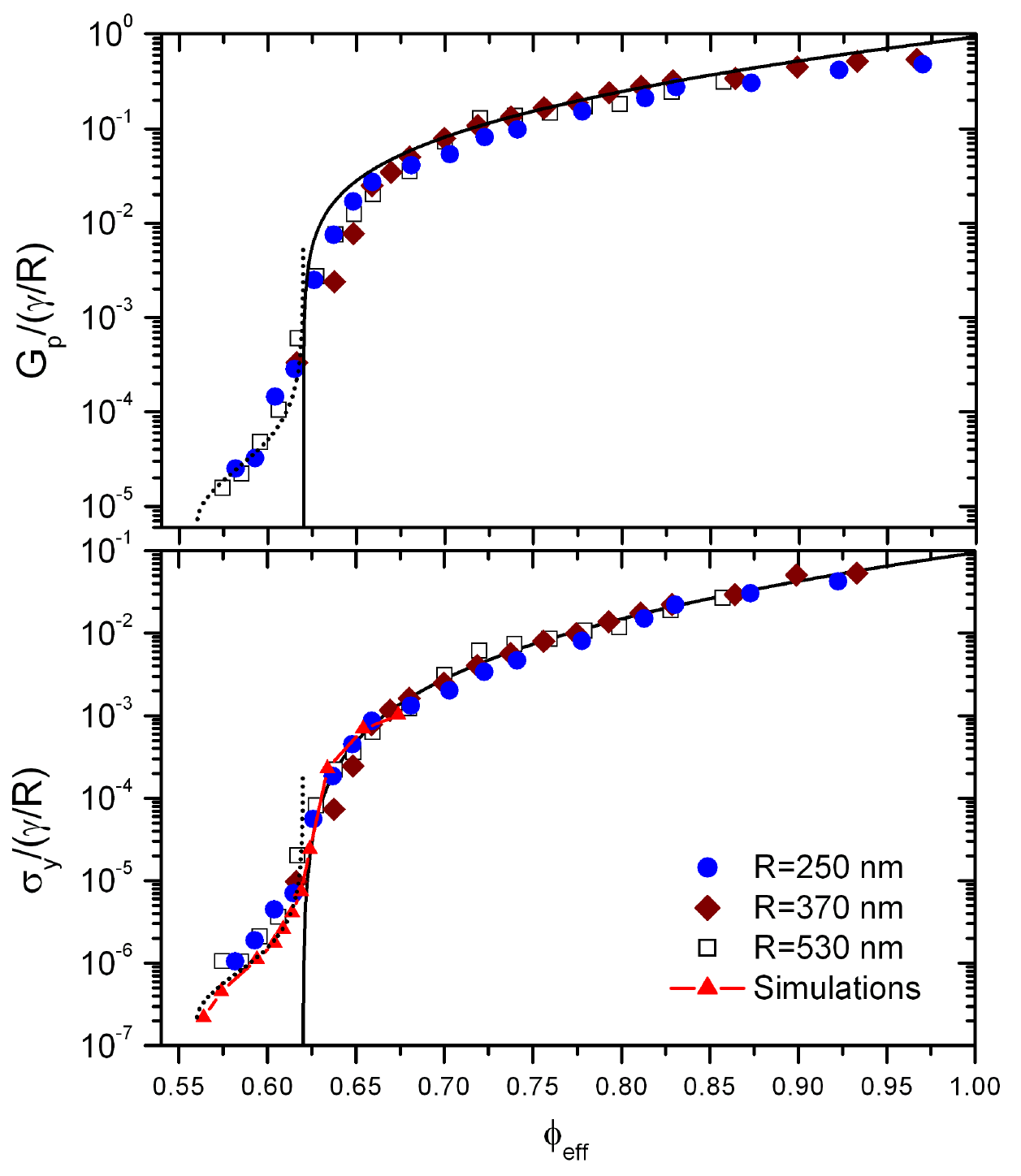}}
\caption{Upper panel: Shear modulus of emulsions with radius $R=250, 370$ and $500$nm. Lower panel: Yield Stress for the same samples. Dotted lines: Scaling predictions for modulus in the glass phase,  Eq.(\ref{scalingglass}) and for  the yield stress with $c_1=0.25,c_2=2.5$ and $G_p=\sigma_y/0.03$. Solid Lines: Theoretical predictions in the jammed phase, Eq.(\ref{yieldstress}) and (\ref{elasticonly}), with $a_1=0.25,a_2=0.05$ and $\phi_J=0.62$, $\phi_g=\phi_J-0.06=0.56$. Solid triangles : simulation results for harmonic spheres with $(k_B T)/\epsilon=10^{-6}$ (data reproduced from Fig. 2(b), ref. \onlinecite{Sollich12}).} \label{datamodyield} 
\end{figure}

An amorphous solid is a state of matter that has both fluid- and solid-like attributes: it has a liquid-like structure, yet mechanically responds to an applied stress as a solid. The conceptually simplest amorphous solid is composed of monodisperse spheres that are non-interacting at a distance yet are repulsive when they make contact  \cite{BookGlassBerthier,LiuReview11,WeeksReview11,BraderReview10}. The repulsive interaction potential at contact can be \emph{soft} or \emph{hard}. Upon rapidly quenching a disordered system above a sphere volume fraction of $\phi\sim 0.6$, a supercooled fluid of spheres enters an amorphous phase, since crystallization is suppressed. Despite decades of intense research on the nature of this transition, it is still not fully understood \cite{Science2005}. In the zero temperature limit, there are no entropic contributions to the free energy and solidity emerges simply from a random packing of spheres approaching what is known as \emph{random close packing} or the \emph{jamming} transition \cite{LiuReview11,Torquato2000}. The location of this transition is well established at $\phi_J \simeq 0.64$ and it plays a key role in the properties of granular spherical packings from the micro- to the macro-scale.
At finite temperatures, however, an additional phase boundary at densities several percentile below jamming, the ergodic-to nonergodic glass transition $\phi_g$, has been conjectured by mode coupling theory (MCT) and as an explanation of light scattering experiments and for hard-sphere systems \cite{WeeksReview11,PuseyvanMegen1986}. In this scenario a nonergodic \emph{glass} is a weak amorphous solid having a density $\phi_g<\phi<\phi_J$ \cite{WeeksReview11,PuseyvanMegen1986,BraderReview10}. 
\newline \indent While many detailed experimental and theoretical studies can be found as $\phi$ approaches $\phi_g$ from the liquid side \cite{WeeksReview11,PuseyvanMegen1986,MasonPRL95,Crassous06}, studies in which $\phi$ crosses both $\phi_g$ and $\phi_J$ are scarce \cite{Petekedis12}. The very existence of a solid glassy phase has been questioned, and it has even been suggested that the glass and jamming transition are the same $\phi_g \equiv \phi_J$ \cite{Brambilla09}. Results obtained from experiments on only one type of experimental system have not yet been able to unravel this question, mainly because it is very difficult to determine exact volume fractions of submicron sized particles \cite{Poon2012}, as a result of non-zero nanoscopic length scales associated with stabilization. Moreover, the very concept of hard-sphere-like systems has been questioned recently \cite{PoonWeeks2013} due the difficulty in designing colloids having a sufficiently sharp repulsive interface. 
\newline \indent In this letter, we study the elastic shear modulus and the yield stress of a colloidal model system of (sub)microscale soft repulsive spheres. We discuss the experimental evidence for a solid glass and a jammed phase crossing the liquid solid transition of dense uniform emulsions. Moreover we demonstrate that applying simple theoretical concepts to these technically relevant systems provides accurate quantitative predictions about their linear and nonlinear mechanical properties.  
\newline \indent An emulsion is a dispersion of droplets of one immiscible liquid in a different liquid \cite{MasonChapter96}. The nano- to micrometer sized droplets of the dispersed phase can be kept almost indefinitely in a metastable state by decorating their interfaces with amphiphilic surfactant molecules. In emulsions, because of the existence of a thin film of continuous phase between droplets, lubrication is present, so solid-solid friction cannot play a role. By contrast to compressible star polymers or microgels \cite{Petekedis12,StarPoly,Crassous06,ScheffoldPRL10}, which do not have volume fractions that are precisely defined, in an emulsion, the droplet volume fraction is well-defined because the liquid within the droplets is incompressible, even if the droplet interfaces can deform under extreme compression. Near and above the liquid-solid transition, droplets in emulsions are not highly deformed and remain nearly spherical. To prevent droplet coalescence, a short-range, screened electrostatic repulsive interaction, provided by an ionic surfactant, is typically present. For the anionic surfactant, sodium dodecyl sulfate (SDS), at $10$ mM concentration in water, the Debye screening length is $\lambda_D \simeq 3.5$nm \cite{BibetteForces}. The distribution of radii $R$ of the droplets can be made to be quite uniform $\Delta R /R \sim 10-12\%$. Herein, we interpret measurements of the linear and non-linear rheology of silicone oil in water near-microscale emulsions stabilized by $10$mM SDS for droplet radii $R=250,370$ and $530$nm. The experimental data that we consider for the elastic modulus $G_p$ and the yield stress $\sigma_y$ have been published previously, and technical details can be found elsewhere \cite{MasonJCIS96,MasonChapter96,WilkingMason2007}. 
\newline \indent To provide perspective, we summarize briefly very recent theoretical predictions for the elastic shear modulus and the yield stress in the glass and in the jamming regimes. Theoretical and numerical work by Ikeda, Berthier and Sollich \cite{Sollich12} addresses the glassy dynamics and jamming of soft spheres. The authors lay out a theoretical scenario about how to distinguish between the mechanical response of a \emph{glassy} system and that of a \emph{jammed} system. Extensive simulations on idealized spheres that have harmonic repulsions beyond contact were carried out in order to map out the transition. What sets this study apart is their explicit focus on describing the physics of soft colloids crossing all relevant regimes. Most previous attempts for model colloidal systems have either concentrated on the approach to the glass transition $\phi \le \phi_g$ \cite{WeeksReview11,PuseyvanMegen1986,MasonPRL95,Crassous06} or on the jamming physics $\phi>\phi_J$ \cite{LiuReview11,CloitreBonncaze2006,Brujic2009}. By considering both, Ikeda \emph{et al.} find that glassy and jammed colloids display qualitatively different characteristics with respect to their mechanical responses. The details of the transition, however, depend sensitively on the nature of particle elasticity (\emph{i.e.} repulsive interaction) when particles touch. For very soft spheres, the transition from the glass to a jammed solid is smeared out and phenomenologically disappears \cite{Yodh2009}. 
\newline \indent The mechanical properties of the glass are entropically driven and scale with temperature and size as $\sim k_B T/R^3$. Ikeda \emph{et al.} suggest the following scaling relation for the yield stress  covering the entire glass phase for hard spheres \cite{Sollich12}
\begin{equation} \frac{{{\sigma _y}}}{{{{{k_B}T} \mathord{\left/
 {\vphantom {{{k_B}T} {{{\left( {2R} \right)}^3}}}} \right.
 \kern-\nulldelimiterspace} {{{\left( {2R} \right)}^3}}}}} \simeq {c_1}\left[ {1 + {c_2}\frac{{{{\left( {\phi  - {\phi _g}} \right)}^{0.7}}}}{{{{\left( {{\phi _J} - \phi } \right)}^{0.8}}}}} \right] \label{scalingglass}
\end{equation} 
with constants $c_1,c_2$ of order unity. MCT calculations for both the yield stress and the modulus are consistent with this scaling, the parameters $c_1,c_2$ however strongly depend on the specific approximations made \cite{FuchsCates2003,FuchsBallauff2005}. The second term in Eq.(\ref{scalingglass}) takes account for the expected divergence of $\sigma_y$ close to the jamming of hard spheres \cite{Brady93,Sollich12}. Since the divergence is avoided for soft spheres, the transition must be somewhat smeared, as is also apparent in simulation results \cite{Sollich12}.
\newline \indent We now turn our attention to the packing fraction dependence of the shear modulus above the jamming threshold $\phi>\phi_J$. In the jamming scenario the modulus is given by a product of the bond strength $k$ and the excess number of contacts \cite{LiuReview11} which has been found to scale as $\Delta Z \propto \sqrt{\phi-\phi_J}$ \cite{LacassePRL96,Mohan2012} and thus
\begin{equation}
G_p = a_1\frac{k}{{\pi R}}\sqrt {\phi  - {\phi _J}} \label{modmodel}
\end{equation}
where $a_1$ is a constant of order unity   \cite{LiuReview11,ScheffoldPRL10,Ciamarra2013}.  Note that $\pi R G_p$ is the spring constant of a particle harmonically bound in a matrix with $G_p$ \cite{HBP}. When applying a sufficiently large stress $\sigma_y$  the sample will yield and subsequently flow. Recent simulations suggest $\sigma_y \sim k (\phi  - {\phi _J})^{1.2}$ \cite{Sollich12,Olsson07} which in turn  provides expressions for the yield stress and the yield strain:

\begin{equation}
{\sigma _y} = {a_2}\frac{k}{{\pi R}}{\left( {\phi  - {\phi _J}} \right)^{1.2}=\left( {{{{a_2}} \mathord{\left/
 {\vphantom {{{a_2}} {{a_1}}}} \right.
 \kern-\nulldelimiterspace} {{a_1}}}} \right){G_p}{\times}{\left( {\phi  - {\phi _J}} \right)^{0.7}}} \label{yieldstress}
\end{equation}
\begin{equation}
\gamma_y\sim \sigma _y/G_p=a_2/a_1 (\phi-\phi_J)^{0.7}  \label{eqyieldstrain} \end{equation}
\begin{figure}[htb]
\centerline{\includegraphics[width=7.5cm]{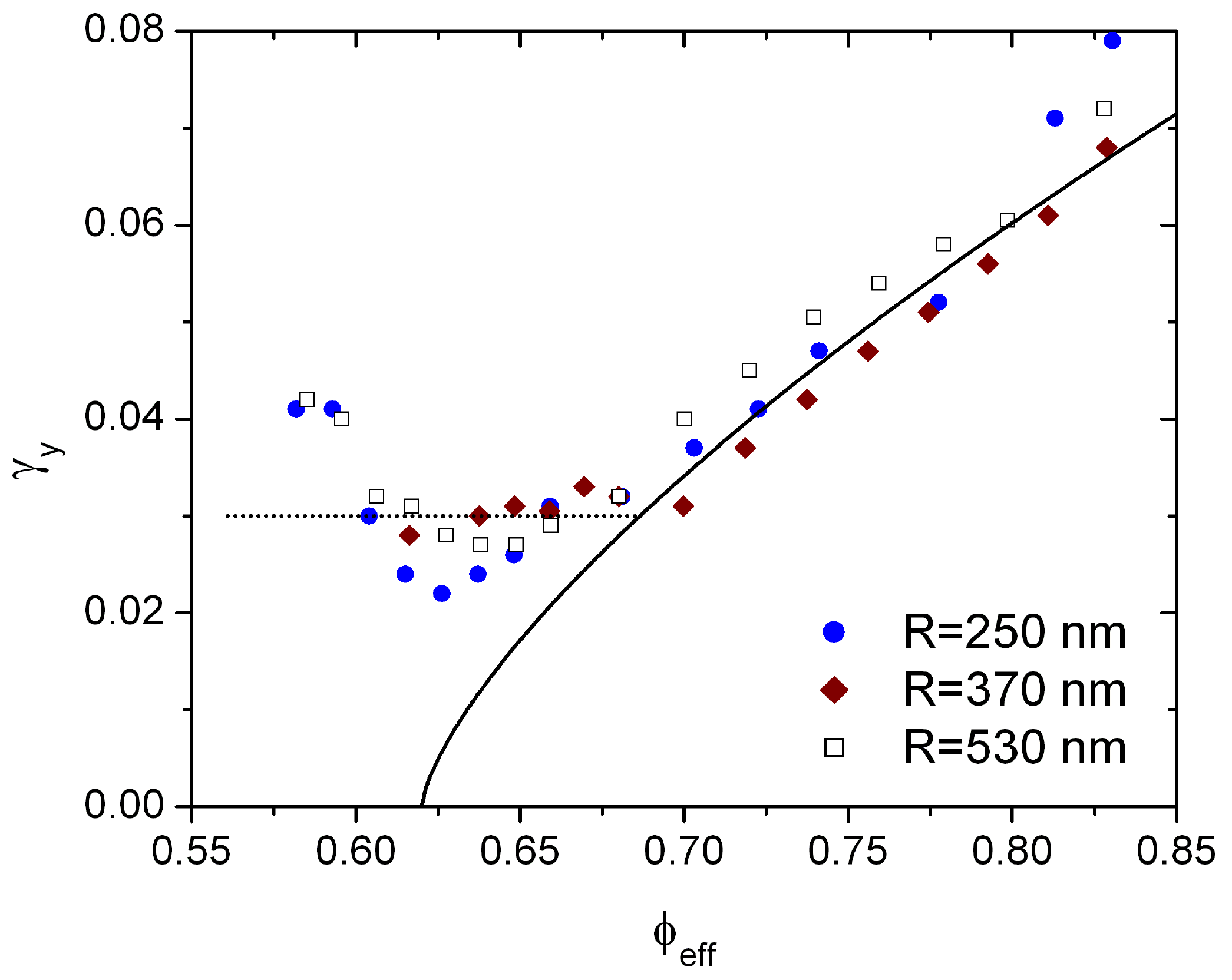}}
\caption{Yield strain $\gamma_y $ as a function of the effective volume fraction. Dotted line : Constant value of 0.03. The solid line shows the prediction from Equation \ref{yieldstress}  with $a_2/a_1=0.2$ and $\phi_J=0.62$.} \label{yieldstrain} 
\end{figure}
\newline \indent We first compare the theoretical predictions with experimental data for the yield strain. The yield strain should be scale invariant and thus the data for different sizes should collapse on a master curve. As shown previously, this is indeed the case for microscale droplets provided the bare droplet volume fraction is converted to an effective volume fraction $\phi_{eff}=\phi \kern 1pt(1+h/2R)^3$, where a $\phi$-dependent effective thickness $h \sim 10 $nm of the interfacial layer, is used to account for screened electrostatic interactions between droplet interfaces \cite{MasonJCIS96, MasonChapter96}.  Moreover the yield strain should be independent of the bond strength according to Eq.(\ref{eqyieldstrain}). Indeed, as shown in Fig. \ref{yieldstrain}, well above $\phi_J$ the yield strain rises and the data is consistent with the scaling predicted by Eq.(\ref{eqyieldstrain}) with $a_2/a_1 \sim 0.2$.
Over $0.60 \leqslant \phi_{eff} \leqslant 0.68$ the yield strain data are approximately constant. At the very lowest measured $0.58 \le \phi_{eff} \le 0.60$, well below $\phi_J$ and very close to measurements of $\phi_g$, $\gamma_y$ is reported to be slightly larger, which would indicate a different scaling of $G_p$ and $\sigma_y$ when approaching the jamming transition (i.e. a different $c_1$ and $c_2$ parameter in Eq.(\ref{scalingglass}), such that the modulus would rise faster than the yield stress approaching $\phi_J$). However this observation could also be due to an increasing uncertainty in the line intersection method on the log-log plot of stress versus strain, which defines the yield point, as the emulsion's low-strain response begins to transition from dominantly elastic to dominantly viscous  \cite{MasonJCIS96}. Given the uncertainty in the experiments and analysis, it is reasonable to say that the yield strain is approximately constant for $\phi_{eff}$ between $\phi_g$ and $\phi_J$, at about $0.030 \pm 0.005$.  The same value for $\sigma_y/G_p$ has been reported by Ballauff, Fuchs and coworkers in a careful study of microgel core-shell particles approaching the glass transition from below \cite{Siebenbuerger2009}. 
\newline \indent We now return to the discussion of the modulus and the yield stress of emulsion. The scaling relations Eq.(\ref{modmodel}) and (\ref{yieldstress}) have been suggested for harmonic spheres having pair-wise interactions described by $V(r)=\epsilon \kern 1pt (1-r/2R)^2$, where the spring constant is $k=\epsilon/2R^2$ \cite{LiuReview11,Ciamarra2013,Sollich12}. We make the assumption that they can also be applied to the case of emulsions but we allow the bond strength $k$ to depend explicitly on the center-to-center distance $r$. The bond strength can be calculated by taking the second derivative of the interaction potential $V(r)$ between two emulsion droplets \cite{LacassePRL96}:
\begin{equation}
V(r) = {3^{ - \alpha }}\epsilon {\left[ {{{\left( {{{2R} \mathord{\left/
 {\vphantom {{2R} r}} \right.
 \kern-\nulldelimiterspace} r}} \right)}^3} - 1} \right]^\alpha } \label{elpotential}  
 \end{equation}
In  the limit of small droplet compression ($r/2R-1\ll1$) the potential reduces to $V(r)=\epsilon (1-r/2R)^\alpha$,  $\epsilon=2 C \gamma R^2 3^\alpha$. Here $\gamma$ is the surface tension. For our case $\gamma=9.8 \cdot 10^{-3}$N/m and the characteristic constants are $C=0.36$, $\alpha=2.32$.  For example for $R=250$nm, $\epsilon \simeq 1.4\times 10^6 k_B T$. We find for $k = {{{\partial ^2}V\left( r \right)} \mathord{\left/
 {\vphantom {{{\partial ^2}V\left( r \right)} {\partial {r^2}}}} \right.
 \kern-\nulldelimiterspace} {\partial {r^2}}}$ :
\begin{equation}
\begin{aligned}
  \frac{{{2{R}^2}k}}{\epsilon} =  & \frac{{6\alpha }}{{{3^{ \alpha }}}}{\left( {\frac{{2R}}{r}} \right)^5}{\left[ {{{\left( {\frac{{2R}}{r}} \right)}^3} - 1} \right]^{\alpha  - 1}} \\ 
   &  + \frac{{{\kern 1pt} 9\alpha \left( {\alpha  - 1} \right)}}{{{2 \cdot 3^{ \alpha }}}}{\left( {\frac{{2R}}{r}} \right)^8}{\left[ {{{\left( {\frac{{2R}}{r}} \right)}^3} - 1} \right]^{\alpha  - 2}} \\ 
\end{aligned} \label{springconstant}
\end{equation}
In a bulk emulsion the interdroplet distance $r$ is set by the droplet number density which in turn is related to the volume fraction occupied by the oil droplets and thus $\phi \propto r^{-3}$. At the jamming transition ($\phi_J\simeq0.64$) the droplets are in direct contact $r=2R$ thus $\phi(r) = \phi_J (2R/r)^3$ and we can write $1-r/2R\simeq1-(\phi_J/\phi)^{1/3}$ \cite{CloitreBonncaze2006}. Combining Eqns.(\ref{springconstant}) and (\ref{modmodel}) provides a simple expression for the shear modulus:
\begin{figure}[htb]
\centerline{\includegraphics[width=7.5cm]{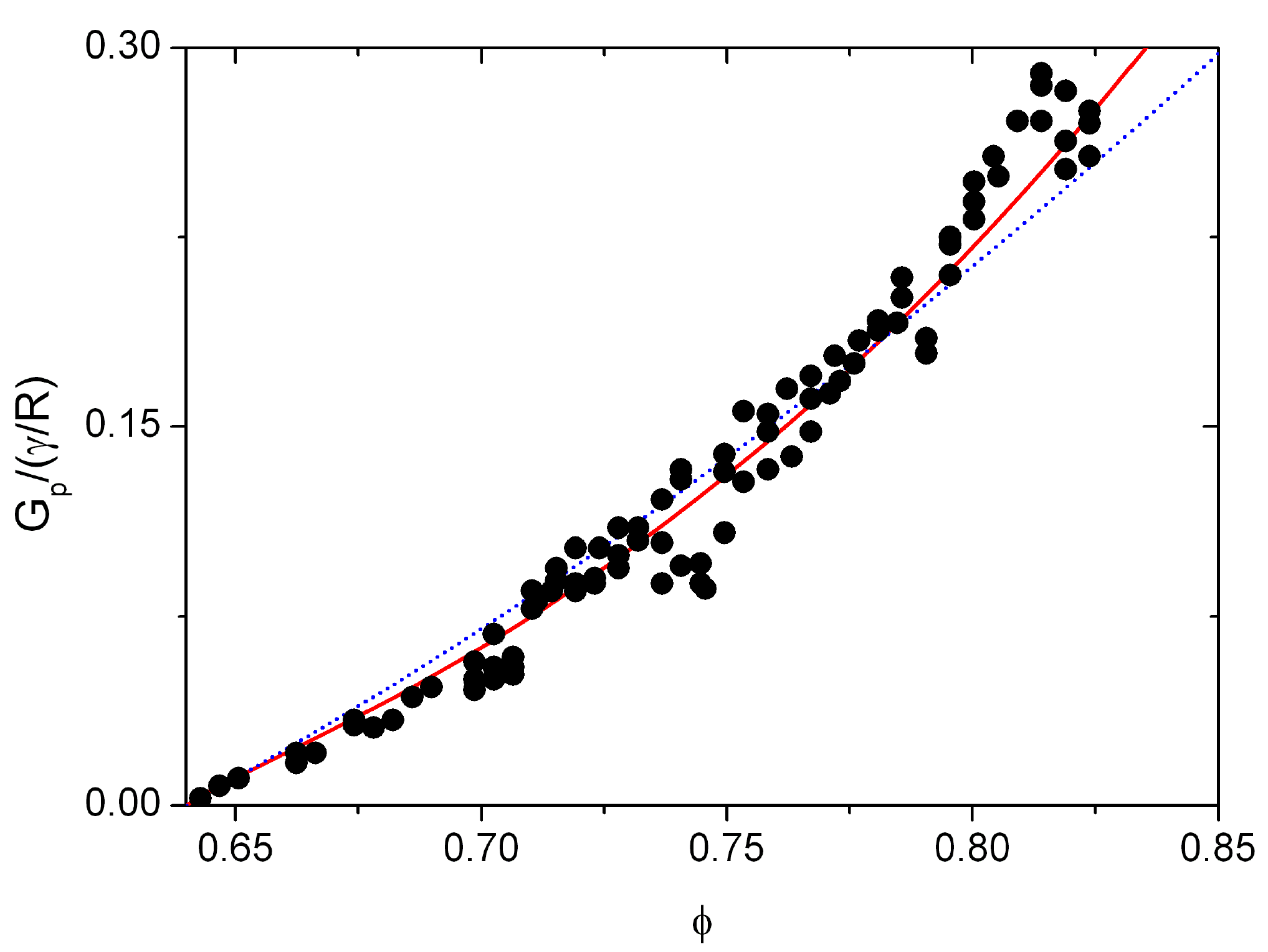}}
\caption{Comparison of  Eq.(\ref{elasticonly}) (solid line, $a_1=0.25$) with numerical simulations of the shear modulus $G_p$ (symbols) \cite{LacassePRL96}. Dotted line: scaling  $G_p\simeq 1.6 \cdot \phi (\phi-\phi_J)$ as reported in \cite{MasonEl}.} \label{LacasseComp} 
\end{figure}
\begin{equation}
{\frac{{{G_p}}}{{{\gamma  \mathord{\left/
 {\vphantom {\gamma  R}} \right.
 \kern-\nulldelimiterspace} R}}}} \simeq 6 a_1  \left[ {{\phi ^{8/3}}{{\left( {\phi  - \phi _J} \right)}^{0.82}} + {\phi ^{5/3}}{{\left( {\phi  - \phi _J} \right)}^{1.82}}} \right]
\label{elasticonly}
\end{equation}
For an estimate of the prefactor $a_1$ in Eq.(\ref{elasticonly}) we compare our model to simulation results for $G_p$ reported by Lacasse et al. \cite{LacassePRL96}. 
For the comparison we have slightly rescaled the $\phi$-values of the simulation data by a numerical factor of $0.978$ to have them extrapolate to the same critical value $\phi_J=0.64$. We find excellent agreement for $a_1=0.25 \pm 0.01$ as shown in the  Fig. \ref{LacasseComp}. This also sets an estimate for the yield stress parameter $a_2 =0.2 a_1\simeq 0.05$. Remarkably, Eq. (\ref{yieldstress})  and Eq. (\ref{elasticonly}) now provide quantitative analytic predictions for the shear modulus and yield stress of emulsions that can be tested directly against experimental data. In Fig. \ref{datamodyield} we compare the predictions in the jammed state, Eq.(\ref{yieldstress}) and (\ref{elasticonly}), with the experimental data for the shear modulus and the yield stress with $\phi_J \simeq 0.62$ as suggested in \cite{MasonJCIS96}. The agreement is remarkable given the difficulty of such a quantitative comparison. 
\newline \indent While other formulas exist for predicting the modulus and yield stress of a disordered uniform emulsion as a function of volume fraction \cite{MasonJCIS96}, they are empirically based and have not, up to now, been derived analytically in a microscopic model that properly incorporates the potential of interaction between droplets and the scaling of the coordination number. Moreover, the scaled $G_p(\phi)$ and $\sigma_y (\phi)$ are self-consistent in this model (i.e. use the same microscopic parameters) and provide good agreement with both linear and nonlinear measurements. Earlier experiments on polydisperse emulsion suggested  $G_p \sim {\phi ^{1/3}}\left( {\phi  - {\phi _J}} \right){\gamma  \mathord{\left/
 {\vphantom {\gamma  R}} \right.
 \kern-\nulldelimiterspace} R}$ \cite{Princen86}. Subsequently it has been shown that the behaviour of uniform emulsion is better described by  the semi-empirical relation ${G_p} \sim \phi \left( {\phi  - {\phi _J}} \right){\gamma  \mathord{\left/
 {\vphantom {\gamma  R}} \right.
 \kern-\nulldelimiterspace} R}$ \cite{MasonEl}. Our result, based on the correct scaling of the excess number of contacts $\Delta Z$, is very close ($\emph{i.e.}$ within about $\pm 10$ percent) to the latter up to densities of about $\phi\sim 0.85$. For $\phi \to 1$ our model extrapolates to $G_p \times ( R / \gamma) \simeq 0.88$ a value slightly larger than reported in \cite{MasonEl}, yet, this value satisfyingly matches a calculation for dry foams and emulsions \cite{Dryfoam}.
\newline \indent Next we address the properties in the glass state and the transition from the glass to the jamming physics, which is governed by the particle softness. Ikeda's measure of softness is given by $T_e=(k_B T)/\epsilon$  \cite{Sollich12}. Computer simulations for harmonic spheres show that for values larger than $10^{-4}$ (very soft spheres) the glassy phase narrows and the transition is smeared out to an extent that in an experiment or simulation glassy physics and jamming cannot be distinguished any more. For our droplets $T_e \sim 10^{-7}-10^{-6}$ which should allow to distinguish both regimes. The glass physics and the onset of jamming should not depend on the details of the interaction potential. We can thus attempt to compare the simulation results for $T_e = 10^{-6}$ reported in \cite{Sollich12} with the  experimental data for the yield stress $\sigma_y$. Note that the numerical data in \cite{Sollich12} is given in units of $\epsilon/(2R)^3$ while our data is normalized by $\gamma / R$. For our emulsion droplets however $(\epsilon/(2R)^3)/(\gamma / R)\simeq 1$ and thus the comparison can be done directly.  To match the jamming transition of the emulsions we slightly shift the numerical data such that $\phi_J=0.62$ in both cases. As shown in Fig. \ref{datamodyield} the agreement is excellent all the way from the glass to the jamming regime. Moreover, comparing the numerical data with Eq. \ref{scalingglass}, we can estimate $c_1 \simeq 0.25$ and $c_2 \simeq 2.5$ using again $G_p \simeq \sigma_y/0.03 $ in the glass (dotted lines in Fig. \ref{datamodyield}). 
\newline \indent Despite the excellent agreement reported here it is worth mentioning that over the range of sizes covered the experimental data does not scale with the droplet size as suggested by entropic origin of the glass elasticity, Eq.(\ref{scalingglass}).  We are unable to discern whether this due to limitations of the theory or due the experimental difficulties in determining such small values with sufficient accuracy or both. To resolve this question additional experiments covering a larger range of droplet sizes will be required in the future. 
\newline \indent Notwithstanding the present work suggests that the glass and jammed phase are well distinguishable for experiments on uniform emulsions. Both the modulus and the yield stress are characterized by a sharp rise close to the jamming transition. This indicates that the smearing of the transition is limited and our emulsion droplets indeed behave as soft spheres with a sufficient stiffness $\epsilon/k_B T$ or bond strength. In reality, emulsion droplets stabilized by an ionic surfactant are slightly more complicated. The presence of the Debye layer will soften the droplet interactions. However, as droplets get into contact, the Debye layer is rapidly compressed and the stiffness increases until it approaches the value $\epsilon/k_B T$ of the elastically coupled droplet core \cite{WilkingMason2007}. Our results suggest that this process takes place over a very limited range of concentrations for the droplet sizes $R\ge250$nm considered here. The situation might be different when considering smaller droplets \cite{WilkingMason2007} or particles having a different interaction potential. 
\newline \indent We thank Joe Brader for illuminating discussions. This work was supported by the Swiss National Science Foundation under grant No. 132736. 
\newline $^\ast$Corresponding author, email: frank.scheffold@unifr.ch

{99}

\end{document}